\documentclass{elsart}
\usepackage{graphicx}
\begin{document}
\title{Centrality and Transverse Momentum Dependence of 
Collective Flow in 158 A GeV Pb+Pb Collisions Measured via Inclusive Photons}
\begin{frontmatter}
\collab{WA98 Collaboration}
\author[inst4]{M.M.~Aggarwal} ,
\author[inst2]{Z.~Ahammed} ,
\author[inst7]{A.L.S.~Angelis} * , 
\author[inst13]{V.~Antonenko} ,
\author[inst6]{V.~Arefiev} ,
\author[inst6]{V.~Astakhov} ,
\author[inst6]{V.~Avdeitchikov} ,
\author[inst16]{T.C.~Awes} ,
\author[inst10]{P.V.K.S.~Baba} ,
\author[inst10]{S.K.~Badyal}
% ...
\address[inst4]{University of Panjab, Chandigarh 160014, India}
\address[inst7]{University of Geneva, CH-1211 Geneva 4,Switzerland}
\address[inst13]{RRC ``Kurchatov Institute'', RU-123182 Moscow, Russia}
\address[inst6]{Joint Institute for Nuclear Research, RU-141980 Dubna, Russia}
\address[inst16]{Oak Ridge National Laboratory, Oak Ridge, Tennessee 37831-6372, USA}
\address[inst10]{University of Jammu, Jammu 180001, India}
% ...
\address[inst2]{Variable Energy Cyclotron Centre,  Calcutta 700 064, India}
% ...
\begin{abstract}
Directed and elliptic flow of
inclusive photons near mid-rapidity in $158 $A GeV Pb+Pb collisions
 has been studied. The data have been obtained with the photon spectrometer
LEDA of the WA98 experiment at the CERN SPS.
 The flow strength has been measured for
various centralities as a function of $p_T$ and rapidity over 
$0.18 < p_T < 1.5 \, \mathrm{GeV}/c$  and $2.3 < y < 2.9$.
The angular anisotropy has been studied relative to an event plane
 obtained in the target
fragmentation region that shows the elliptic flow to be in-plane. 
The elliptic flow has also been studied using two-particle correlations
and shown to give similar results.
A small directed flow component is observed.
Both the directed and elliptic flow strengths increase with $p_T$.
The photon flow results are used to estimate the
 corresponding neutral pion flow.
\end{abstract}
\end{frontmatter}

\section{Introduction}     

Heavy ion collisions at relativistic energies 
provide a means to
study the properties of nuclear matter at high temperature and density.
In such collisions it is expected that a high density interaction 
zone is formed. If this system thermalizes, the thermal pressure
will generate collective transverse expansion \cite{Hoff94}.
Such collective flow, and especially its anisotropy,
 will reflect the time evolution of the pressure 
gradients of the system and can provide information on the 
equation of state (EOS) in the initial phase \cite{Olli92,Sorg97}
and during the expansion \cite{Stoc86}, and in particular about
the possible formation of the Quark Gluon Plasma (QGP)
\cite{Hung95,Risc96}.

QGP formation requires high energy density and local thermalisation of the
system.  While transverse energy measurements indicate
 the attainment of high energy
densities \cite{EDENS}, the degree of thermalization has not been 
unambiguously determined.  Measurements of collective flow may be one of the
strongest hints related to the degree of thermalisation.  One may consider two
extreme cases: the low density limit, where the mean free path is comparable
or larger than the system size for which cascade models may be appropriate,
 and the hydrodynamic limit, where the
mean free path of the particles is much less than the system size.  The study
of collective flow as a function of beam energy and system size, transverse and
longitudinal momentum, and particle species may allow to separate these two
scenarios and to recognize the hadronic or partonic
 (QGP) nature of the reaction
\cite{Volo00}.

The anisotropic flow of charged fragments
 has been measured in nuclear collisions at
various beam energies: at 0.1--1.0 A GeV
\cite{PballC,KaosCo,FopiCo}, at $10 $A GeV
\cite{E877Co}, at $158 $A GeV \cite{NA49Co,WA80Co,WA93Co,WA98Co,CERESCo} 
and also at $\sqrt{s_{NN}} = 130$ and $200$ GeV
\cite{PhenCo,STARCo,PHOBOSCo}. 
The first $p_T$-integrated photon flow measurement
was performed by the WA93 experiment \cite{WA93Co}
for 200 A GeV S+Au collisions, where photons
were measured with the Photon Multiplicity Detector (PMD).
In the present work we report first results of photon collective 
flow measurements in 158 A GeV Pb+Pb collisions using photons identified and 
momentum analyzed in the lead glass 
calorimeter LEDA of the WA98 experiment. Preliminary results have been 
presented in~\cite{Niko03,Bath03}.

\section{Experimental Setup}

The data presented here were obtained in the WA98 experiment 
\cite{misc:wa98:proposal:91} for
158 A GeV Pb+Pb collisions at the  CERN SPS. The WA98 setup consisted of 
large acceptance hadron and photon spectrometers, calorimeters for
forward and transverse energy measurements, and detectors for
photon and charged particle multiplicity measurements.

 The centrality of the event was determined by the total
transverse energy, $E_T$, measured with the 
mid-rapidity calorimeter
(MIRAC)~\cite{Awes89}, which covered the pseudo-rapidity
 range of $3.5 < \eta < 5.5$. It was placed at 24.7 meters
 downstream from the target and consisted of 30 stacks, each
divided vertically into 6 towers, of size 20 x 20 cm$^2$ each.
 The MIRAC measured both the transverse
electromagnetic $E^{em}_T$ and hadronic $E^{had}_T$ energies.
Events with large $E_T$
correspond to the most central collisions with small impact
parameter.
The minimum bias trigger required a beam trigger with a MIRAC
transverse energy greater than a low threshold.

The Plastic Ball spectrometer, used for the reaction plane (RP)
determination, had full azimuthal coverage in the pseudorapidity
range of $-1.7 < \eta < 0.5$ (i.e. in the target fragmentation region with
polar angles $70^{\circ} < \theta < 160^{\circ}$).
It consisted of 655 detector modules and
allowed to identify pions, protons, deutrons, tritons, $^3$He, and $^4$He
with kinetic energies of 50 to 250 MeV by the $\Delta E - E$
method. Each module comprised a slow 4 mm thick CaF$_2$ $\Delta E$
scintillator followed by a fast plastic scintillator, both read out by a
common photomultiplier \cite{Bade82}. 

Photons, of which $\sim 85 \%$  
originate from $\pi^0$ decay \cite{Direct}, were detected in the
electromagnetic calorimeter LEDA, a highly segmented 
photon detector located 22.1 m downstream of
 the target and covering the photon
rapidity region $2.3 < y < 2.9$, i.e. backwards of mid-rapidity
 ($y=2.9$).
LEDA consisted of 10080 TF1 lead-glass modules,
read out by FEU-84 photomultipliers.
The photomultiplier high voltage was generated
on-base with custom developed  \cite{ref:ne95} Cockcroft-Walton 
voltage-multiplier type bases which were
individually controlled by a VME processor.
The photomultiplier signals were digitized with a custom-built ADC system
\cite{ref:wi94}.
The dimensions of each 
module were 4x4x40 cm$^{3}$ (14.3 radiation lengths depth and 1.1
Moliere radius width). Each group of 24 modules had its own calibration and
gain monitoring system based on a set of 3 LEDs mounted inside
a sealed reflecting cover dome. Each module viewed the reflected
LED light through an aperture on the front surface, while the LED 
light was simultaneously monitored by a PIN-photodiode \cite{ref:pe96}.

LEDA was calibrated with 10 GeV electrons in the X1 beam at the
CERN SPS in the years 1993-1994. Electron beams with energies from
3 GeV to 20 GeV were used to measure the energy
and position resolution, and the energy non-linearity.
The measured energy resolution was 
$\sigma / E = (5.5 \pm 0.6)\% / \sqrt{E} + (0.8 \pm 0.2)\%$
and the measured position resolution was 
$\sigma / E = (8.35 \pm 0.25) \mathrm{mm} / \sqrt{E}
 + (0.15 \pm 0.07) \mathrm{mm}$.
A more detailed description of the WA98 setup
is given in \cite{WA98Co,misc:wa98:proposal:91,Direct}.

\section{The Methods}

Two complementary methods have been used in this analysis of anisotropic flow: 
the study of single-particle angular distributions with respect to an 
estimated reaction plane (\textit{reaction plane method})
and the study of two-particle correlations (\textit{correlation method}). 
Both methods assume that the underlying azimuthal distribution of particles
with respect to the reaction plane (RP), which is the plane
 that contains the impact 
parameter and beam direction vectors, can be described by a
 Fourier decomposition:
\begin{equation}
  \frac{1}{N} \frac{dN}{d(\phi - \Psi_{RP})}
 = 1+2v_{1} \cos \left(\phi - \Psi_{RP} \right)
+2v_{2} \cos \left( 2(\phi - \Psi_{RP}) \right),
\end{equation}
where $\phi$ is the azimuthal angle of the emitted particle, and
$\Psi_{RP}$ is the azimuthal angle of the RP. 
The anisotropic flow is characterized by 
the values of the Fourier coefficients $v_1$, for directed flow, and
$v_2$, for elliptic flow. 

\subsection{The Reaction Plane Method}

The conventional reaction plane method \cite{Volo96,Posk98} uses the 
distribution of particles in their azimuthal angle relative to 
the estimated reaction plane.
Because the true reaction plane is not known in the experiment, 
one has to establish an \textit{event plane} (EP) from the measured particles 
as an estimate for the reaction plane.  
 
Particles measured in the target fragmentation region show significant
directed flow -- proton, deuterons, and heavier fragments 
are emitted in the reaction plane
in one direction, while pions are emitted in the opposite direction, 
as has been measured with the Plastic Ball detector~\cite{Peit99}. 
This information has been used to calculate an event plane angle from the 
particles measured in the Plastic Ball:
\begin{equation}
\Phi_{EP}=
\tan^{-1} \left (\frac{\sum_{i=1}^N E_T^{i} \sin \phi_i}
{\sum_{i=1}^N E_T^{i} \cos \phi_i} \right),
\end{equation}
where the sum runs over all fragments and identified positive pions.
$E_{T}^{i}$ and $\phi_i$ are
the transverse kinetic energy and azimuthal angle in the laboratory frame
of the $i-$th particle, respectively. For pions,  $\phi$ was replaced
by $\phi + \pi$  to account for the opposite sign pion directed flow~\cite{Peit99}.

The azimuthal distribution of photons detected by the 
LEDA calorimeter has been studied
relative to this event plane angle. The distributions are studied as a function
of $\Delta \Phi = \phi_{\gamma} - \Phi_{EP}$ and are fitted with:
\begin{equation}
  \frac{1}{N} \frac{dN}{d\Delta \Phi} = 
1+2v_{1}^{obs} \cos \left( \Delta \Phi \right)
+2v_{2}^{obs} \cos \left( 2 \Delta \Phi \right).
\label{eq:fit1}
\end{equation}

The measured EP doesn't coincide exactly 
with the true RP because of  the finite number of detected particles
and resulting fluctuations.
Because of this finite reaction plane resolution, the 
coefficients obtained from the
fits have to be corrected by dividing them by the event plane resolution
correction factors (RCF$_n$):
\begin{equation}
  v_n = \frac{v_n^{obs} } {RCF_n}. 
\end{equation}
 
The resolution correction functions RCF$_n$ are given by~\cite{Posk98}:
\begin{eqnarray}
    RCF_n  & = & \langle \cos (n (\Phi_{EP} - \Psi_{RP})) \rangle \nonumber \\
    & = & 
  \frac {\sqrt{\pi}} {2 \sqrt{2}} \chi_m \exp 
\left( \frac {-\chi_m^2} {4} \right)
   \left[ I_{\frac{k-1}{2}} \left( \frac {\chi_m^2} {4} \right) + 
    I_{\frac{k+1}{2}} \left( \frac {\chi_m^2} {4} \right) \right] 
    \label{eq:cf}   
\end{eqnarray}
where $\Psi_{RP}$ is the true RP angle, $\chi_m$ is the resolution parameter,
which is proportional to the square root of the multiplicity, 
$m$ is the order of the Fourier component used for calculation
of the event plane and $k=n/m$. 

Since $\Psi_{RP}$ is unknown,
the RCF's must be determined from the measured EP themselves.
This can be done by a subevent analysis in which each event, 
in this case consisting of hits in the Plastic Ball detector, is randomly
divided into two subevents (A and B) and for each subevent the EP
angle  $\Phi_{EP}$ ($\Phi_{A}$ or $\Phi_{B}$)  is calculated. 
The quantity  $ \langle \cos(n (\Phi_{A} - \Phi_{B}) )\rangle $ 
is determined directly from the subevent correlation
function. It is then used in  equation~\ref{eq:cf} to obtain the 
parameter $\chi_m^{sub}$ for the subevent multiplicity,
which is then used to calculate $\chi_m=\sqrt{2} \chi_m^{sub}$
for the full event, and finally
to obtain $RCF_n$. 
In this analysis we have 
used $m=1$ ($k=1$) for $RCF_1$ and for $RCF_2$ ($k=2$). 

\subsection{The Correlation Method}

Alternatively, the flow values have been obtained from the
azimuthal correlations of photon pairs as described in 
\cite{Olli92,Posk98,Wang91,Lace93,Lace01}. 
The correlation function is calculated as:
\begin{equation}
C_{\gamma \gamma} (\Delta \phi) \equiv \frac{d^2N/d\phi_1 d\phi_2}
{dN/d\phi_1 \cdot dN/d\phi_2},
\end{equation}
where $\Delta \phi\ = \phi_1 - \phi_2$. The $C_{\gamma \gamma}$
are calculated as the ratio of the true two-photon distribution to 
the pair distribution from mixed events. This is necessary to correct for 
distortions from the limited acceptance of the photon detector.
In the event-mixing procedure we have taken care to mix only events with 
similar global properties and to use identical cuts, especially regarding 
the two-cluster separation within the detector.
The two-photon correlation functions have then been fitted with:
\begin{equation}
C_{\gamma \gamma} (\Delta \phi) = 1+2v_{1}^{2} \cos(\Delta \phi)
+2v_{2}^{2} \cos(2(\Delta \phi)).
\label{eq:fit2}
\end{equation}
From these fits, no significant $v_1$ component was observed. Since 
very little directed flow is expected near mid-rapidity, we have 
set $v_1 \equiv 0$ in this analysis. 

If collective flow is dominant, this method should be equivalent
to the previous one since (a)
the correlation between every particle and the RP induces a
correlation amongst the particles, and (b) correlating two
subevents amounts to summing two-particle correlations
\cite{Borg01}. The correlation 
method has the advantage that no EP determination, 
and therefore no resolution correction, is needed. However, non-flow 
correlations, such as back-to-back correlations due to momentum 
conservation~\cite{Bath03}, should be taken into account. 

\section{Analysis}

The data presented here were taken in 1995 and 1996 
with the 158 A GeV Pb ion beams of the CERN SPS. 
Pb targets of 495 and 239 mg/cm$^2$ thickness were used. 
About $10^7$ events were analyzed.
The events were divided into 8 centrality classes 
defined by intervals in the total $E_T$ measured by MIRAC,
as summarized in 
Table~\ref{tab:cent}. 
The centralities are expressed
as fractions of the minimum bias cross sections as a function
of the total $E_T$, measured by MIRAC. 
In addition, a multiplicity of greater than 3 
fragments measured in the Plastic Ball was demanded to allow for a 
reasonable determination of the EP. This has a significant effect 
for the peripheral bins and causes a slight bias towards higher 
multiplicity within the bin.  

In addition to the 8 classes of centrality shown in 
Table~\ref{tab:cent}, 
studies were performed with combined centrality classes:
2~+~3 (47--83\%), 4~+~5~+~6 (13--47\%) and
7~+~8  (0--13\%).
The table also shows the number of participants $N_{part}$
as calculated in a Glauber type calculation discussed in \cite{wa98:scaling}. 

To further suppress the hadron contamination in the photon sample,
only those showers in the calorimeter have been used which satisfy
 the following cuts:
\begin{itemize}
\item The measured energy was greater than 0.75~GeV. This cut suppressed 
minimum ionizing particles.
\item The lateral dispersion of the shower was less than a maximum value.
 This suppressed
showering hadrons.
\end{itemize}
These cuts kept the hadron
contamination in the photon sample to less than $\approx 7\%$ 
\cite{Direct,Berg92}. 
 
\begin{table}
\centering
\begin{tabular}{|c|c|c|c|}

\hline
  \multicolumn{2}{|c|}{ $E_T$ class} & $E_T $ (GeV) & $N_{part}$   \\
\hline
\hline
 1& $ 83 - 100 \%$ & $0 - 28.35 $ & 10 $\pm$2 \\
\hline
 2& $ 65 - 83 \%$ & $28.35 - 79.05$ &28 $\pm$2 \\ 
\hline
 3& $ 47 - 65 \%$ & $79.05 - 161.55$ &63 $\pm$2 \\
\hline
 4& $ 24 - 47 \%$ &  $161.55 - 281.05$ &133 $\pm$3 \\
\hline
 5& $ 19 - 24 \%$ & $281.05 - 318.05$ &205 $\pm$2 \\ 
\hline
 6& $ 13 - 19 \%$ & $318.05 - 361.55$ & 247 $\pm$2 \\
\hline
 7& $  6.5 - 13 \%$ & $361.55 - 410.95$ & 291 $\pm$2 \\
\hline
 8& $  0 - 6.5 \%$ &  $ > 410.95$ & 351 $\pm$1 \\
\hline

\end{tabular}
\caption{Centrality classes used in this analysis. The  
percentage of the measured minimum bias cross section 
included in each class is given. Also given is the corresponding average
number of participants  for each class
with an estimate of the systematic error. Cuts on $E_T$ are for 1995 data set, 1996 data set cuts are shown in~\cite{wa98:pions}.}
\label{tab:cent}
\end{table}

The observed raw $\Phi_{EP}$-distributions showed a 
variation due to detector biases, such as dead channels and 
inefficiency, of less than $5\%$. 
This non-uniformity has to be removed before extracting 
the flow strength from the measured correlation functions. This 
has been done by two methods:
\begin{itemize}
\item The real distribution was divided by the equivalent distribution 
for mixed events, where $\phi_{\gamma}$ and $\Phi_{EP}$
are taken from different events.
\item In accumulating the distributions, the entries were weighted 
with the inverse of the $\Phi_{EP}$ distribution.
\end{itemize}
Both methods gave consistent results within errors, we have used 
the second method for the final result.

\begin{figure}[ht]
\begin{center}
\includegraphics{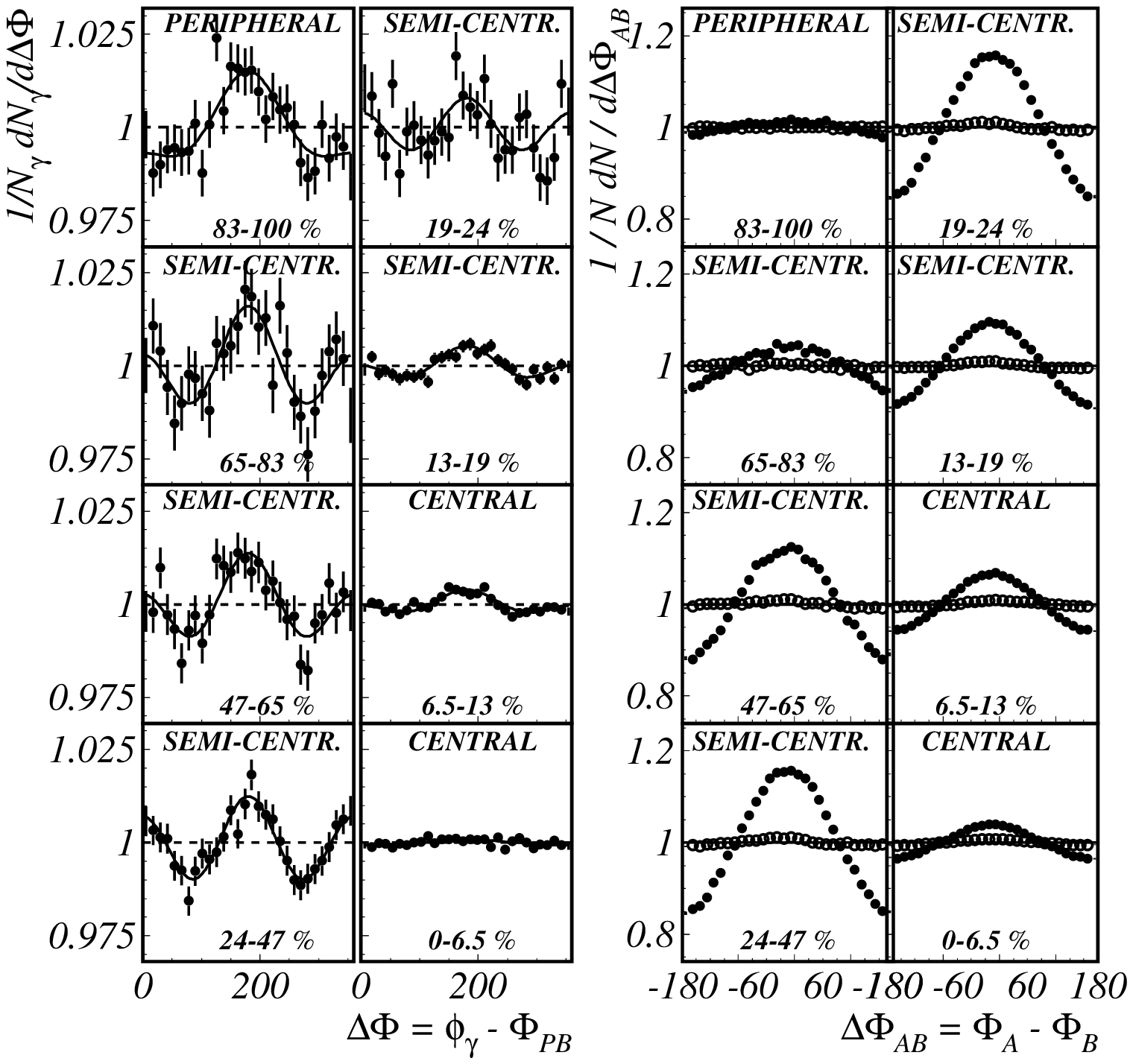}
\caption{ \small Left: The measured photon azimuthal 
 correlation functions for different centralities. The solid lines show fits
 to equation~\ref{eq:fit1}. 
 Right: Subevent correlation functions of particles measured in the 
 Plastic Ball for different centralities. The filled circles show the 
 measured correlation functions.
 The open circles show the results 
for mixed subevents.
\protect\label{fig1}}
\end{center}
\end{figure}

Fig.~\ref{fig1} (left) shows examples of the  measured photon azimuthal
correlation functions for different centralities. A clear modulation
is seen, especially in the semi-central classes, which can be 
described well with fits to equation~\ref{eq:fit1}.

The resolution correction factors RCF's for each centrality
were determined as described above from
the subevent correlation functions 
as shown in Fig.~\ref{fig1} (right). In this case 
the Plastic Ball acceptance correction was done by using mixed 
subevents, i.e. subevents from different events. 
The deviation of the mixed subevents
from unity, i.e. the order of magnitude of this correction, 
was less than $2\%$ (as shown in Fig.~\ref{fig1} (right)).

\begin{figure}[t]
\begin{center}
\includegraphics{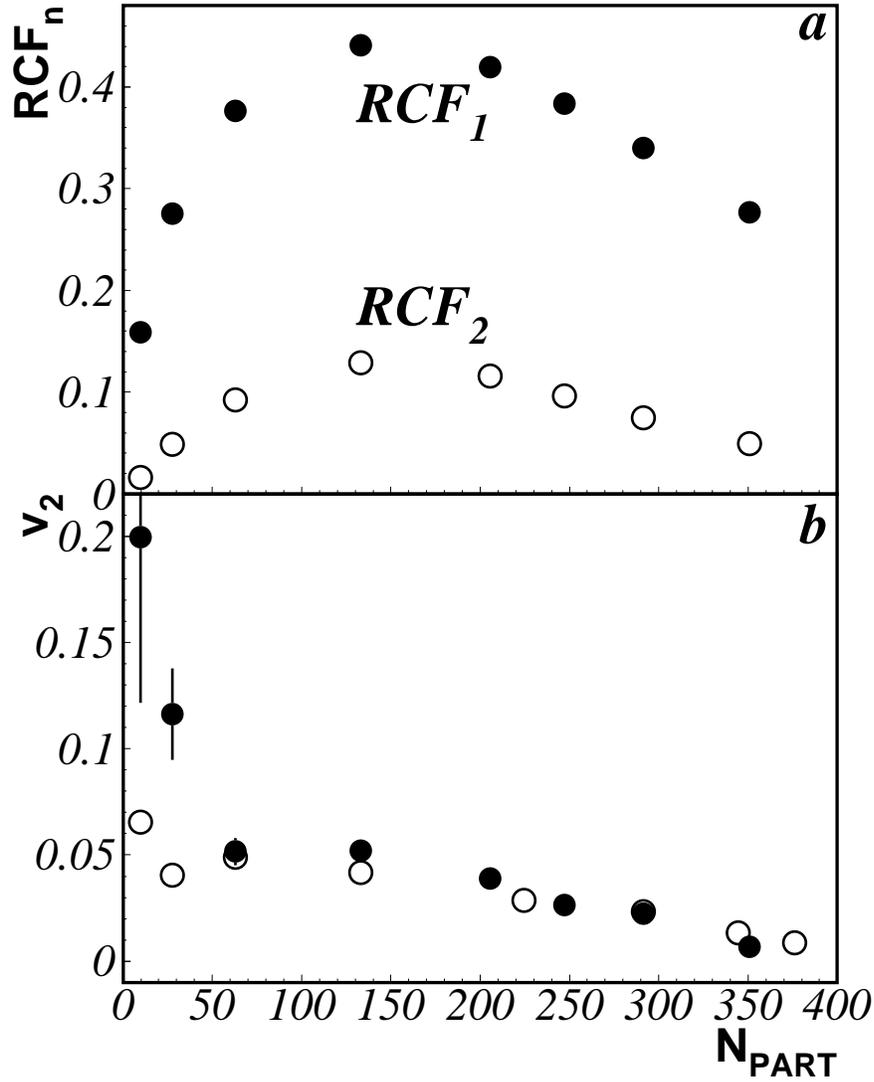}
\caption{\small (a) The resolution correction factors
 obtained from subevent
 correlation functions of particles measured in the 
target fragmentation region
 as a function of the number of participants. 
 The filled circles show the values for directed flow,
 the open circles for
elliptic flow. 
(b)  The elliptic flow $v_2$ of photons extracted by the reaction 
 plane method 
 (solid circles, $p_T \ge 0.18 $ GeV/$c$)
  and by the correlation method (open circles,
  $p_{T1} \ge 0.18 $ GeV/$c$, $p_{T2} \ge 0.18 $ GeV/$c$),
  integrated over
 $p_T > 0.18 $ GeV$/c$ and $y = $2.3--2.9
 as a function of the number of participants. 
 \protect\label{fig2}}
\end{center}
\end{figure}

Fig.~\ref{fig2}a  shows the values of the resolution 
correction factors RCF's
determined in this way as a function of the number of participants.
A stronger subevent correlation implies a better determination of the RP,
and the corresponding values of the RCF's
are larger \cite{Dani85}. It is seen, that the subevent correlation
 is strongest
for semi-central events, whereas for peripheral and central
events the quality of the EP - determination is worse.
The RCF's  shown in Fig.~\ref{fig2}a
 have been obtained for the 1996 beam time.
The RCF's have been calculated and applied independently
 for the 1995 beam time, with
values found to be smaller by $\sim 20 - 40 \%$.

\begin{figure}[ht]
\begin{center}
\includegraphics{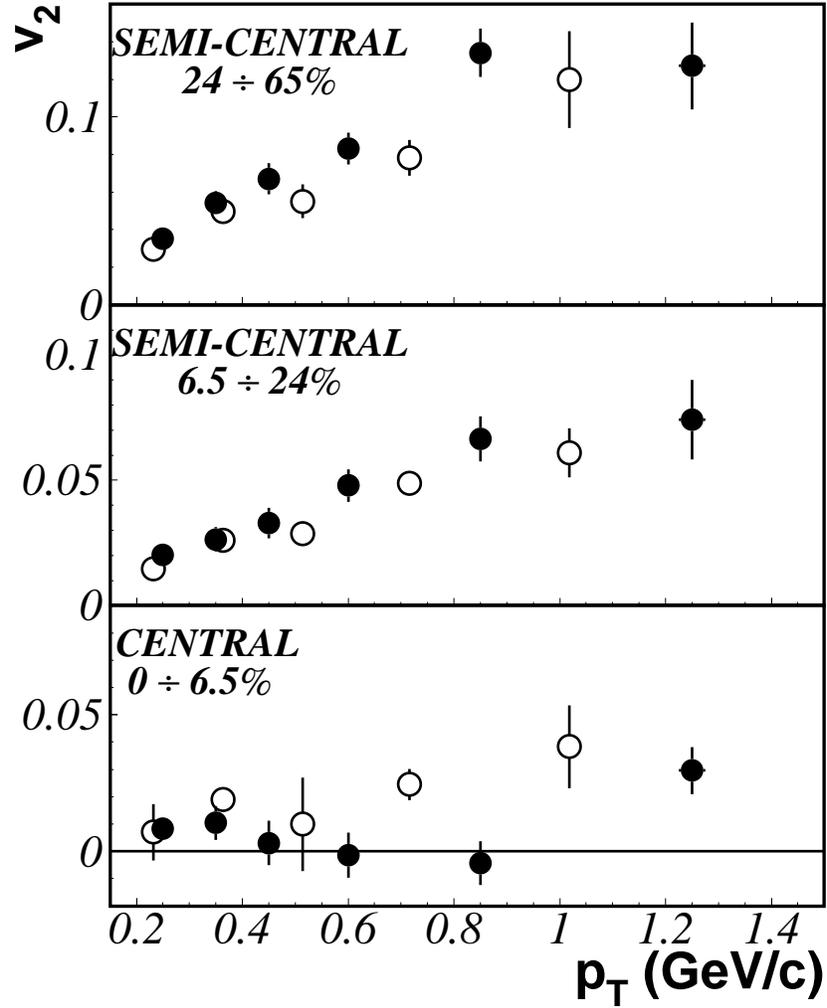}
\caption{\small The elliptic flow $v_2$ of photons extracted by the reaction 
plane method 
(solid circles, $p_T \ge 0.18 $ GeV/$c$)
 and by the correlation method
 (open circles, $p_{T1} \ge 0.18 $ GeV/$c$, $p_{T2} \ge 0.18 $ GeV/$c$),
integrated over $y = $2.3--2.9 for three centrality selections.
 \protect\label{fig3}}
\end{center}
\end{figure}

The  $v_2$ elliptic flow values for photons have also been
obtained from the two-photon correlation functions for 
various centrality classes, although not all of these precisely match
those used for the reaction plane method analysis~\cite{Bues02}.
The $v_2$ elliptic flow values for photons obtained by the two
methods are compared in Figs.~\ref{fig2}b and \ref{fig3}.
The two photon correlation functions have been fitted with 
Eq.~\ref{eq:fit2} to extract $v_2$ with $v_1 \equiv 0$.
Fig.~\ref{fig2}b shows
the centrality dependence of $v_2$ integrated over $p_T > 0.18$ GeV$/c$. 
The solid circles are the
 values obtained from the
reaction plane method and the open circles are those from the
 correlation method.  
For centralities
corresponding to $N_{part} > 50$ there is very good agreement
 between the two methods.
Fig.~\ref{fig3}  shows the elliptic flow as a function of $p_T$
for central and two semi-central classes.
There is good agreement between the methods.

One should note that the most peripheral classes suffer from
 several problems: The 
determination of the reaction plane has a large uncertainty,
 especially fluctuations due 
to the low multiplicity play a role here. In addition, for the
 correlation method it has 
been seen that there may be considerable non-flow components for
 these centralities
\cite{Bath03} which would influence the extraction of the flow.
 Furthermore, we have 
neglected directed flow in the correlation method which may also 
lead to systematic 
errors in the $v_2$ determination. The results from the reaction 
plane method are used in the discussions that follow.

As a consistency check we have performed the reaction plane analysis
 independently
on the data from the two different beam periods, where the 
quality of the RP determination
was found to be different. The resulting flow values are in good agreement. 

\section{Systematic Error}

The systematic errors of the obtained coefficients include:
\begin{itemize}

\item uncertainties in the event plane determination in the Plastic Ball
        due to non-uniformity of acceptance and efficiency or 
        imperfect particle identification, estimated as $< 2\%$.

\item uncertainties in the photon angular distributions due to
charge particle contamination of photons in LEDA. This contamination
is less than $\approx 7\%$ (\cite{Direct}). The uncertainties are
related to a difference in the observed flow between charged pions 
and pion decay photons and were estimated as $< 3.5\%$.
\end{itemize}

  Other sources of the systematic errors
have been investigated by comparing the results obtained
under different conditions.  The following checks have been performed:

\begin{enumerate}

\item Different weights ($E_T$ vs. $p_T$) have been used for the 
        determination of the event plane.

\item Different acceptance regions of the Plastic Ball detector
         ($70^{\circ}<\theta<160^{\circ}$ or  
      $60^{\circ}<\theta<160^{\circ}$) have been used.
      
\item Different identification cuts for 
        the Plastic Ball fragments were used.

\item The event plane was determined with and without including pions
 in the Plastic Ball.

\item Results of the two different beam periods were analyzed separately.

\end{enumerate}

Another possible contribution to the systematic error is due to
 non-flow correlations.
Among such non-flow effects relevant for azimuthal
correlations are the correlations due to momentum conservation,
long- and short-range two- and many-particle correlations
(due to quantum statistics, resonances, jet or mini-jet
 production, etc.). The contribution of non-flow correlations 
scales as $1/N$, where $N$ is the multiplicity of particles
used to determine the event plane.
The ``momentum conservation'' contribution
increases with the fraction of particles detected,
and the relative contribution of Bose-Einstein
correlations would be independent of $N$.
The effect of non-flow correlations in the reaction plane method is expected 
to be small as the event plane is determined from particle which have a 
large pseudorapidity separation $\Delta \eta > 1.8$ from the photons. 
The non-flow contributions to the event plane determination have been
 investigated
by studying the dependence of the RCF's on the Plastic Ball multiplicity. 
The conventional subevent analysis was performed excluding various fractions of
the Plastic Ball particles to investigate the deviation
of the flow parameter $\chi_m$ 
from $\chi_m \approx \sqrt{N}$ (see eq.~\ref{eq:cf}).

All of these systematic checks lead to a total systematic error estimate 
of less than  $\pm 17\%$ for both measured values
of the flow coefficients.
Unless explicitly stated otherwise,
the systematic errors are not included in the figures.

\section{Results}

\begin{figure}[ht]
\begin{center}
\includegraphics{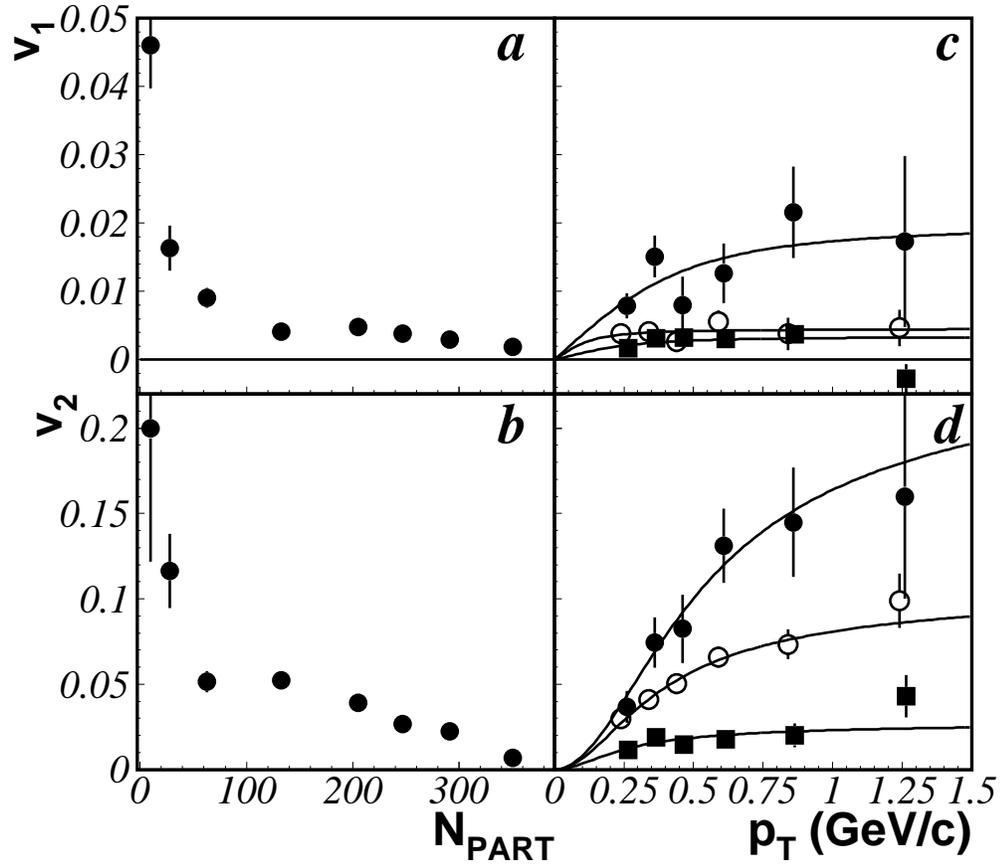}
\caption{\small Directed flow ($v_1$) (a) and 
elliptic flow ($v_2$) (b) of photons
integrated over
$p_T > 0.18 $ GeV$/c$ and $y = 2.3 - 2.9$
as a function of the number of participants. 
Directed flow ($v_1$) (c)
 and elliptic flow ($v_2$) (d) of photons
integrated over $y = $2.3--2.9
as a function of $p_T$ for various centralities.
Solid circles show results for semi-central events (47--83\%),
open circles show results for semi-central events (13--47\%),
solid squares show results for central events (0--13\%).
Solid lines are fitted results to a blast wave model functional form.
 \protect\label{fig4}}
\end{center}
\end{figure}

Fig.~\ref{fig4}a,b shows the final values of the directed flow $v_1$ and the 
elliptic flow $v_2$ coefficients as a function of the number of participants.
The data were integrated over 
$p_T > 0.18 \,\mathrm{GeV}/c$ and over rapidity $y = 2.3 - 2.9$.

Both types of flow decrease in strength with the number of participants.
The sign of all values is positive with the sign convention that positive 
$v_1$ corresponds to the directed flow direction of the protons in the 
projectile fragmentation direction and positive $v_2$ corresponds to 
the in-plane direction. Thus, the directed flow of photons 
below mid-rapidity is in the same direction as the directed flow of 
pions, oppositee to the protons, 
in the target fragmentation region, and the elliptic flow is 
oriented in the reaction plane.

The transverse momentum dependence
of the photon flow integrated
over $y = 2.3 - 2.9$ is shown in Fig.~\ref{fig4}c,d
 for different centrality classes.
Both flow coefficients show an increase with $p_T$ which is
 compatible with a
blast wave behavior (see below,\cite{STARCo},
\cite{NA4903}). 

To be able to compare the flow of photons to other experimental results 
we have attempted to establish the relation between
 the observed photon flow and 
the underlying flow of the parent neutral pions. This has been investigated 
in Monte-Carlo simulations. In the simulations,
$\pi^0$'s were generated according to the $p_T$  spectrum
  measured by WA98 \cite{Direct}.
The $\pi^0$ azimuthal distributions were modulated with
 directed and elliptic flow components.
The values of the  $\pi^0$ $v_1$ and $v_2$ and their
$p_T$-dependence have been constrained by the photon measurement.
The $p_T$-dependence of the azimuthal asymmetry has
been parameterized following a
simple hydrodynamically motivated blast wave model, described in 
 \cite{STARCo},\cite{Huov01}, and generalized in \cite{NA4903}
 to also describe $v_1$ :
\begin{eqnarray}
v_n(p_T) = \frac{\int\limits_0^{2\pi}d\phi_b \cos(n\phi_b)I_n(\alpha)K_1(\beta)[1+2s_n\cos(n\phi_b)]}{\int\limits_0^{2\pi}d\phi_b I_0(\alpha)K_1(\beta)[1+2s_n\cos(n\phi_b)]}
\label{flow_pt}
\end{eqnarray}
where the harmonic \textit{n} can be either 1 or 2,
$I_0,I_n$, and $K_1$ are the modified Bessel functions, 
$\phi_b=\phi-\Psi_{RP}$, 
$\alpha(\phi_b)=(p_T/T_f)\sinh[\rho (\phi_b)]$, 
$\beta(\phi_b)=(m_T/T_f)\cosh[\rho(\phi_b)]$,
$s_n$ is the surface emission parameter,
and $T_f$ is the freeze-out temperature.
The azimuthal flow rapidity is given as 
$\rho(\phi_b)=\rho_0+\rho_a \cos(n\phi_b)$ with
 $\rho_0$
is the mean transverse expansion rapidity
 ($v_0=\tanh[\rho_0]$)
and $\rho_a$ is the amplitude of its
 azimuthal variation, respectively. 
Further details are given in \cite{NA4903}.

Without regard to the physical interpretation of the parameters 
of this model it can be used to provide a convenient parameterization
of the $p_T$ dependence of the flow. For this purpose Eq.~\ref{flow_pt}
can be further simplified by setting $\rho_a=0$. The integrals
can be solved to reduce the expression to two parameters, 
$a_n$ and $b_n$. The expressions for the $p_T$ dependence of the 
directed and elliptic flow are then:
\begin{eqnarray}
v_1(p_T) = a_1 \frac{I_1(b_1 p_T)}{I_0(b_1 p_T)}
\label{v1_fit}
\end{eqnarray}
\begin{eqnarray}
v_2(p_T) = a_2 (1 - 2\frac{I_1(b_2 p_T)}{b_2 p_T I_0(b_2 p_T)})
\label{v2_fit}
\end{eqnarray}
The measured photon $v_n(p_T)$ are described well by these
expressions (see Fig.~\ref{fig4}c,d).
The extracted $a_n$ and $b_n$ were constrained to have a
smooth centrality dependence.

The simulated $\pi^0$ $p_T$ distributions have been parameterized as:
\begin{eqnarray}
\frac{1}{N_{Event}}\frac{d^2N}{dydp_T}=\frac{1}{N_{Event}}E\frac{d^3N}{dp^3}p_T=C^\prime p_T \left( \frac{p^\prime_0}{p^\prime_0+p_T} \right)^{n^\prime}
\label{Hagedorn}
\end{eqnarray}
with parameters $C^\prime, p^\prime_0,n^\prime$ taken from the measured
$\pi^0$ results (\cite{Direct}). Jetset 7.4 was used to generate the 
$\pi^0$-decay photons which were then filtered in the simulations
with the detector acceptance and efficiency.

\begin{table}
\centering
\begin{tabular}{|l|c|c|c|c|c|c|c|c|}

\hline
& \multicolumn{4}{|c|}{Fit to $\gamma$} & 
\multicolumn{4}{|c|}{Extracted for $\pi^0$} \\

\hline

Centrality & $a_1$ & $b_1$ & $a_2$ & $b_2$ & $a_1$ & $b_1$ & $a_2$ & $b_2$ \\

\hline
\hline
 
 1 & 0.008 & 3.06 & 0.255 & 5.3 & 0.005 & 3.19 & 0.335 & 3.046 \\
  
\hline

 2 & 0.0075 & 3.06 & 0.24 & 5.25 & 0.0075 & 3.06 & 0.29 & 3.217 \\
  
\hline

 3 & 0.0072 & 3.06 & 0.22 & 5.2 & 0.0072 & 3.06 & 0.28 & 3.09 \\
  
\hline

 4 & 0.006 & 3.06 & 0.17 & 5.1 & 0.0078 & 3.06 & 0.217 & 2.98 \\
  
\hline

 5 & 0.005 & 3.06 & 0.12 & 5.0 & 0.005 & 3.256 & 0.158 & 2.87 \\
  
\hline

 6 & 0.0045 & 3.06 & 0.095 & 4.9 & 0.0044 & 3.06 & 0.118 & 2.93 \\
  
\hline

 7 & 0.004 & 3.06 & 0.06 & 4.8 & 0.0042 & 3.06 & 0.068 & 3.1 \\
  
\hline

 8 & 0.003 & 3.06 & 0.03 & 4.7 & 0.0026 & 3.06 & 0.028 & 3.8 \\
  
\hline

\end{tabular}
\caption{
The blast wave parameters $a_n$ and $b_n$ of
 Eqs.~\ref{v1_fit} and \ref{v2_fit} for various centralities
obtained by fits to the WA98 inclusive
photon data (left) and the corresponding parameters extracted
for neutral pions (right) from simulations that reproduce the 
inclusive photon results.}
\label{tab:blast}
\end{table}

In the simulations, the  $\pi^0$'s were generated with an azimuthal 
asymmetry with $p_T$ dependent $v_n$ parameterized by
Eqs.~\ref{v1_fit} and ~\ref{v2_fit}.
 As an initial ansatz, the measured photon
$v_n = g(p_T)$ for each centrality was used for the $v_n = f(p_T)$ of the
simulated pions. The
output $v_n = g'(p_T)$ of the simulated decay photons was compared
to the measured photon dependence and used to adjust the $\pi^0$ 
$v_n$ for the next iteration. The procedure was iterated until
the simulated and measured photon results were in agreement, which 
typically required two to three iterations.
 The $a_n$ and $b_n$
coefficients obtained from the fit to the photon results and the coefficients 
extracted for the 
$\pi^0$'s by the iteration procedure are summarized in 
Table~\ref{tab:blast}. 
 The ratio of the flow coefficients
extracted for the simulated decay photons to the input pion flow coefficients
provided the correction factors $k_n(p_T) = v_n^\gamma(p_T)/ v_n^\pi(p_T)$
used to extract the $\pi^0$ $v_n$ from the measured photon $v_n$
for $p_T > 0 $ GeV$/c$.
The systematic errors on the $k_n$ were determined
from the uncertainties $\Delta$ of the fit parameters: 
$a_n \pm \Delta a_n$ , $b_n \pm \Delta b_n$,
$p^\prime_0 \pm \Delta p^\prime_0$ and 
$n^\prime \pm \Delta n^\prime$. Within errors,  the final results
were found to be consistent with an analysis in which a simple
linear  $p_T$ dependence of the $v_n^\pi(p_T)$ was assumed.

\begin{figure}[ht]
\begin{center}
\includegraphics{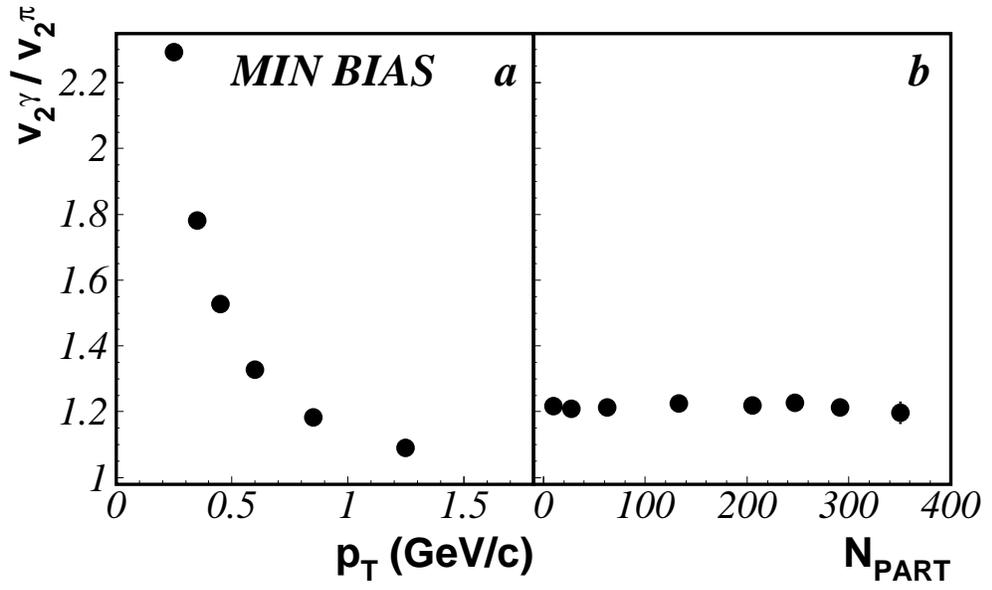}
\caption{\small  The ratio of photon elliptic flow to
 parent pion elliptic flow coefficients.
(a) $k_2 = v_{2}^{\gamma}/v_{2}^{\pi}$
 for min. bias,
(b) $k_2 = v_{2}^{\gamma}/v_{2}^{\pi}$
 for different centrality bins.
 \protect\label{fig5}}
\end{center}
\end{figure}

Fig.~\ref{fig5}a,b shows $k_2$, the ratio of the flow coefficients
of photons and pions, obtained from the simulations.
The $k_1$ ratio is found to be independent of centrality and $p_T$
with a value of $\langle k_1 \rangle = 1.075 \pm 0.003$ with a 
systematic error  less than $\pm 12 \%$.
Since $v_1^\pi$ is small, $k_1 \approx 1$.
The $k_2$ ratios are also found to be independent of centrality
with $\langle k_2 \rangle = 1.215 \pm 0.002$ and a systematic error less than 
$\pm 8 \%$, but strongly dependent on $p_T$.
The observation that the $k_n$ ratios are greater than unity can 
be understood as a simple effect of the $\pi^0$ decay and the
fact that the $v_n$ increase with $p_T$. 
Comparing photons and $\pi^0$'s at a given $p_T$, the 
decay photons will have been produced from 
$\pi^0$'s with larger $p_T$ and hence larger $v_n$, giving $k_n>1$
(see Fig.~\ref{fig5}a and b).
The $k_n$ results shown in Fig.~\ref{fig5}a and b have been used 
to estimate the neutral pion flow values from the measured
photon flow.

\begin{figure}[ht]
\begin{center}
\includegraphics{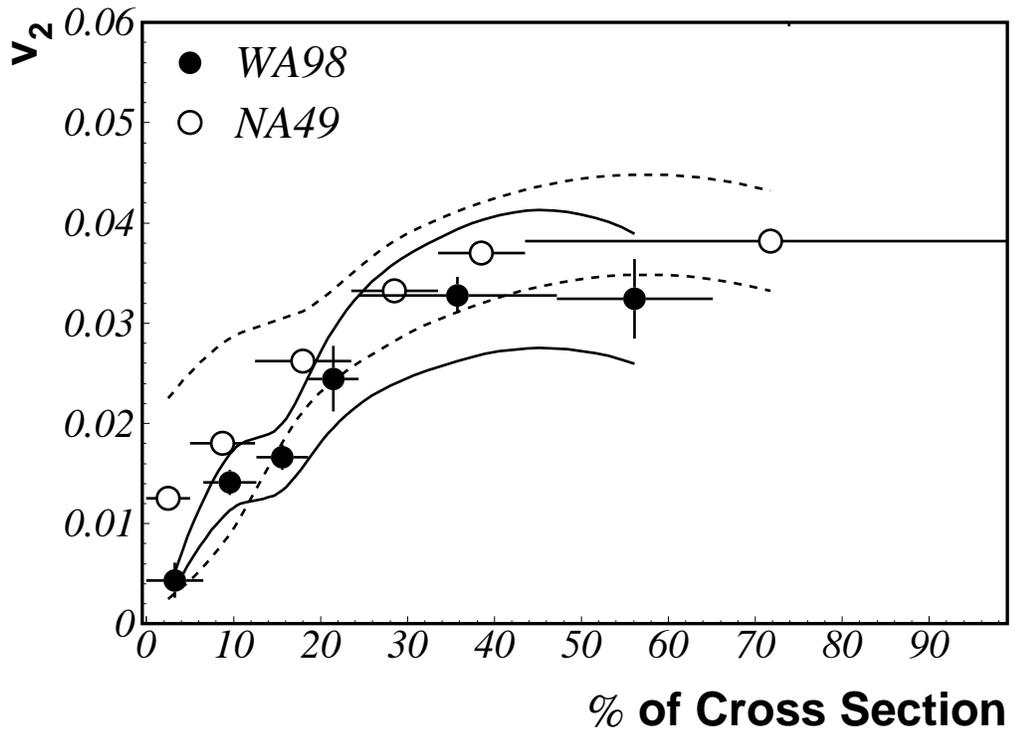}
\caption{\small The centrality dependence of pion
elliptic flow $v_2$ extracted from 
the measured photon flow from WA98 (solid circles,
$-0.6 <y_{CM}< 0 $), 
compared to pion flow results of NA49 
(open circles, $0 <y_{CM}< 2.1 $,~\cite{NA4903}).
 \protect\label{fig6}}
\end{center}
\end{figure}

The $\pi^0$ $v_2$ extracted from the measured WA98 photon $v_2$
as a function of centrality 
are compared to the results from NA49 for charged pions
\cite{NA4903} in Fig.~\ref{fig6}. 
For this comparison
the WA98 results for photons with $p_T > 0.18 $ GeV$/c$ were corrected
to the expectation for the pions without $p_T$ threshold 
($p_T > 0 $ GeV$/c$) as for NA49 results with the correction 
factor $k_1^0=1.30\pm0.036$ and $k_2^0=1.59\pm 0.003$. The
systematic errors on the WA98 points including the errors of the 
measured photon $v_n$ values and the additional error   
of the $k_n^0$ correction are less than $\pm20\%$, except for
the lowest $p_T$ points where the upper systematic errors increase
(Fig.~\ref{fig8}). 
The systematic errors
on the NA49 data points
vary from $13 \%$ for the most peripheral to 
$80 \%$ for the most central bin. The two measurements are
seen to be in agreement.

\begin{figure}[ht]
\begin{center}
\includegraphics{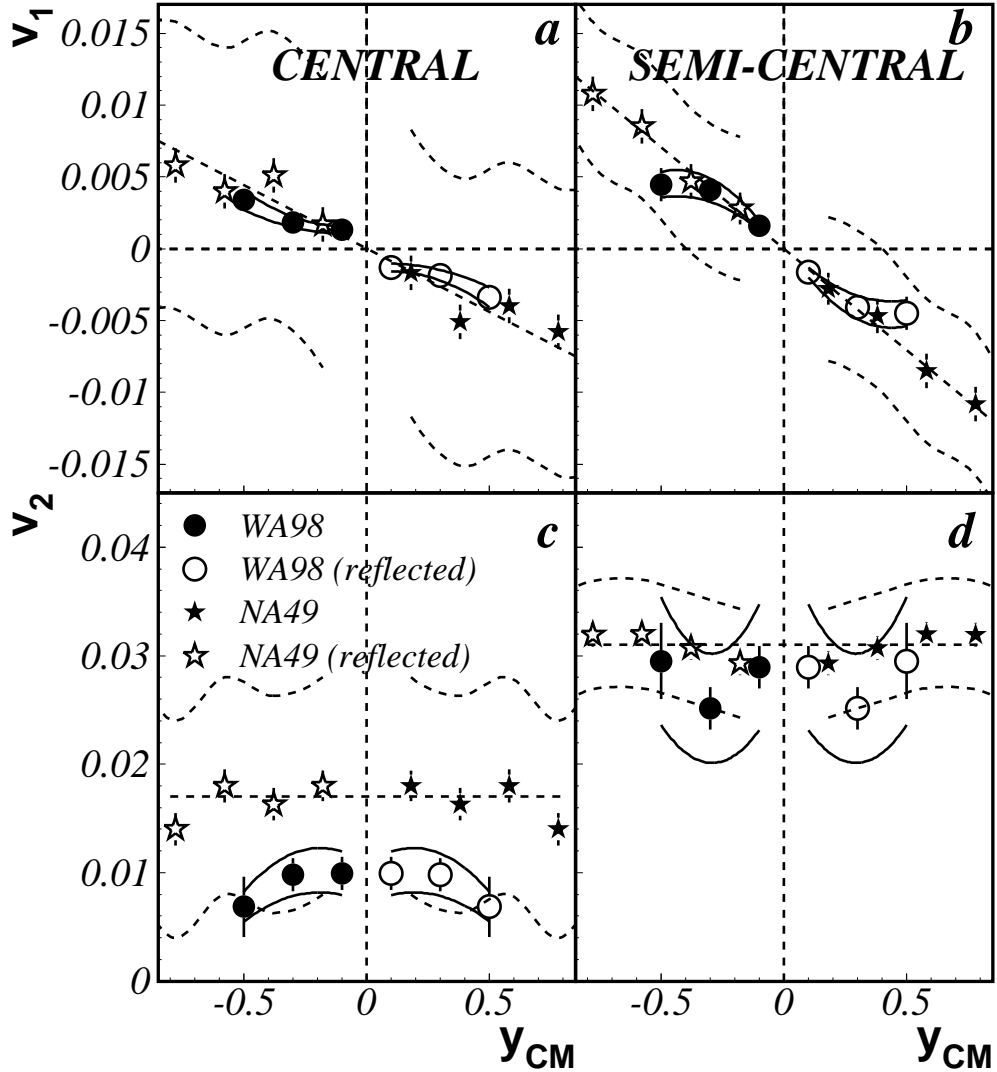}
\caption{\small 
 The rapidity dependence of 
pion flow extracted from 
the measured photon flow from WA98
(circles)
compared to pion flow results of NA49 
(stars,~\cite{NA4903}).
 The results are shown
for central on the left panel 
(WA98: 0--13\%, NA49: 0--12.5\%) and
semi-central on the right panel
(WA98: 13--47\%, NA49: 12.5--33.5\%) selections
for the directed flow (a,b) and elliptic flow (c,d).
Systematic error bands are shown on (c) for NA49 (dashed curves)
and for WA98 (solid curves).
Open points are reflected at mid-rapidity.
The dashed lines indicating the rapidity dependence
 are only meant to guide the eye.
\protect\label{fig7}}
\end{center}
\end{figure}

\begin{figure}[ht]
\begin{center}
\includegraphics{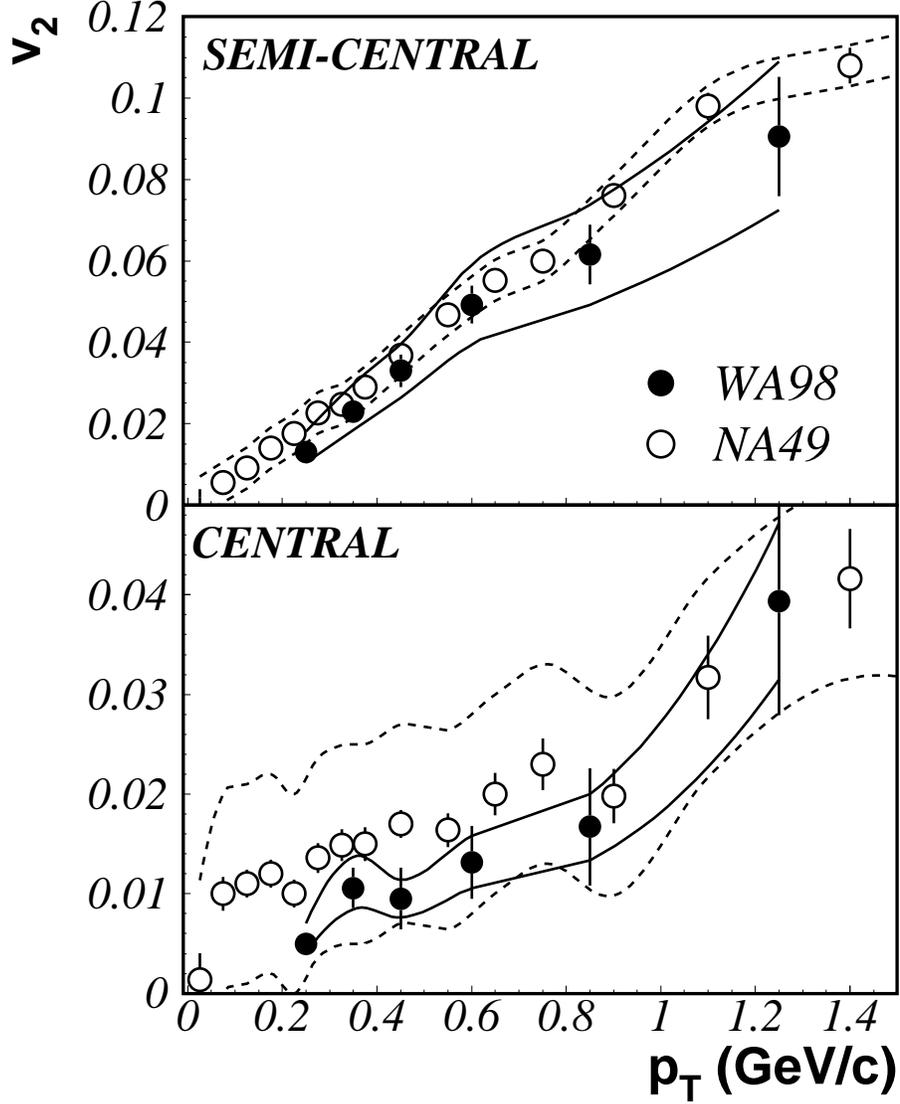}
\caption{\small 
 The transverse momentum dependence of the 
$\pi^0$ elliptic flow extracted from 
the measured photon flow from WA98
(solid circles)
compared to $\pi^+$ elliptic flow results of NA49 
(open circles,~\cite{NA4903}).
 The results are shown for
semi-central selections
(top: WA98: 13--47\%, $-0.6 < y_{CM} < 0$,
  NA49: 12.5--33.5\%, $0 < y_{CM} < 0.8$)
and for central selections  
(Bottom: WA98: 0--13\%, $-0.6 < y_{CM} < 0$,
 NA49: 0--12.5\%, $0 < y_{CM} < 0.8$).
Systematic error bands are shown for NA49 (dashed curves)
and for WA98 (solid curves).
 \protect\label{fig8}}
\end{center}
\end{figure}

The rapidity dependence of the pion flow coefficients is
shown in Fig.~\ref{fig7}
 for two centrality selections. The WA98
photon $v_n$ with $p_T > 0.18 $ GeV$/c$ were corrected
to the expectation for $\pi^0$'s without $p_T$ threshold using 
the correction factors $k_n^0$ given above. 
The NA49 results for pions with
$p_T > 0 $ GeV$/c$ with appropriate corrections (\cite{NA4903}) 
are also shown and seen to be in agreement.
Similar agreement between the two experiments 
is seen  in the transverse momentum dependence of the $v_2$
shown in Fig.~\ref{fig8}.

\begin{figure}[ht]
\begin{center}
\includegraphics{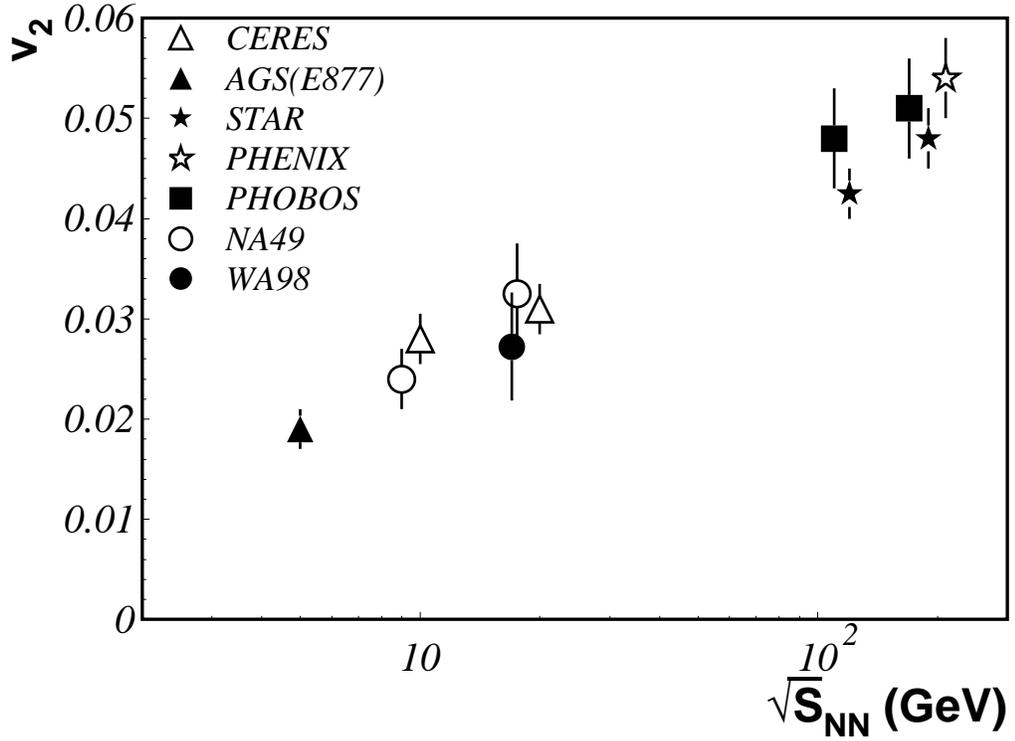}
\caption{\small 
Comparison of the WA98 $p_T$-integrated
elliptic flow results at $\sqrt{s_{NN}}=17$ GeV 
(13--47\%, $-0.6 < y_{CM} < 0$)
with results from other experiments for different 
collision energies.
The WA98 result is the $\pi^0$ elliptic flow extracted from the
measured photon elliptic flow, as described in the text. 
The total statistical and systematic error is shown on the WA98 point. 
 \protect\label{fig9}}
\end{center}
\end{figure}

Fig.~\ref{fig9} shows the WA98 $\pi^0$ $v_2$ coefficient
deduced from the measured photon $v_2$ (corrected by $k_2^0$)
together with a compilation of results~\cite{NA4903}
from other experiments. Results are shown from E877~\cite{Volo00,E877Co}, 
CERES~\cite{CERES}, NA49~\cite{NA4903}, PHOBOS~\cite{PHOBOSv2}, 
PHENIX~\cite{PHENIXv2}, and STAR~\cite{STARv2}, 
The WA98 result follows the general trend 
of the smooth increase of the elliptic flow 
with increasing beam energy.
 It should be kept in mind that the $p_T$ and rapidity coverages
as well as centrality selections differ for the various 
experiments.

Since the strength of the elliptic flow should be proportional to the 
initial eccentricity of the collision zone, it is useful to normalize 
the measured $v_2$ to the eccentricity of the reaction geometry when
investigating the systematics of the elliptical flow~\cite{Sorg99}. 
The initial spatial eccentricity was calculated within 
a Glauber model calculation~\cite{wa98:scaling} as:

$$
\epsilon=\frac{<y^2>-<x^2>}{<y^2>+<x^2>}
$$

where x and y are the participant nucleon 
coordinates in the plane perpendicular to the beam
and x denotes the in-plane direction.

It is of interest to plot the quantity $v_2/\epsilon$ versus the
particle density as estimated by the $dN/dy$ of charged particles divided
by the area of the overlap region S~\cite{Volo00,NA4903}. 
Neglecting the weak incident energy and centrality dependence of the 
average $p_T$,  the scaled particle density is proportional to the initial
energy density in the Bjorken estimate~\cite{Bj}. 
Fig.10 shows the WA98 result for 
$v_2/\epsilon$ for neutral pions together with results for charged 
pions from NA49 \cite{NA4903}
and results for charged particles from STAR \cite{LRay03,Adle02} 
and E877 \cite{Volo00}. The RHIC
data have been corrected for their $p_T$ cutoff, the errors are statistical.
For the WA98 points the $dN/dy$
are taken from WA98 measurements~\cite{wa98:scaling}.

The WA98 results confirm the previous observation~\cite{Volo00,NA4903} of
a universal dependence of  $v_2/\epsilon$ on the particle density. 
It should be noted that results at different incident energies but similar
average particle density correspond to dramatically different collision
geometries. Generally, a central collision at low energy will have a
particle density similar to a more peripheral collision at higher energy.
Thus, the universal dependence demonstrates that the 
scaled $v_2$ depends only on the initial energy density, 
or initial pressure, rather than being dependent on the particle density times
pathlength, which would be the expectation if only partial thermalization  
was attained. 
The observed increase of the scaled elliptic flow with increasing
particle density is qualitatively similar to that seen in a recent
hydrodynamical model study~\cite{Tean04} in which the $v_2$ value for
fixed impact parameter increases smoothly with increasing $dN_{ch}/dy$
corresponding to the increase in pressure.
That systematic study
of hadron spectra and flow results indicated that existing data are best
reproduced by hydrodynamic model 
calculations with an equation of state which include a transition to 
a Quark Gluon Plasma Phase at a critical temperature of $T_c=165$ MeV with a
latent heat of 800 MeV~\cite{Tean04}.

\section{Summary}

Directed and elliptic flow of photons have been measured 
in 158 A GeV Pb+Pb collisions 
using the LEDA electromagnetic calorimeter
and the Plastic Ball detector of the WA98 
experiment. The conventional reaction plane method
for directed and elliptic flow 
and the pair-correlation method for elliptic flow have been used
in the analysis. The elliptic flow values obtained by the two
methods are consistent.

The centrality and $p_T$ dependences of the photon
directed and elliptic flow coefficients were presented. In-plane
elliptic flow is observed. Both flow values increase
for more peripheral collisions and with increasing $p_T$.

For comparison with charged pion measurements, the neutral pion
flow coefficients have been
extracted from the measured photon flow coefficients 
using Monte Carlo simulations.
The Monte Carlo simulations demonstrate that the average photon 
flow coefficients are $\approx$7--21\% greater
than the parent $\pi^0$ flow coefficients.
The extracted neutral pion flow results 
are compatible with the NA49 charged pion flow and with
the general trend of the elliptic flow behaviour
as a function of beam energy. The universal dependence of
the elliptic flow, scaled by the eccentricity of the initial
nuclear overlap, on particle density is confirmed.

\begin{figure}[ht]
\begin{center}
\includegraphics{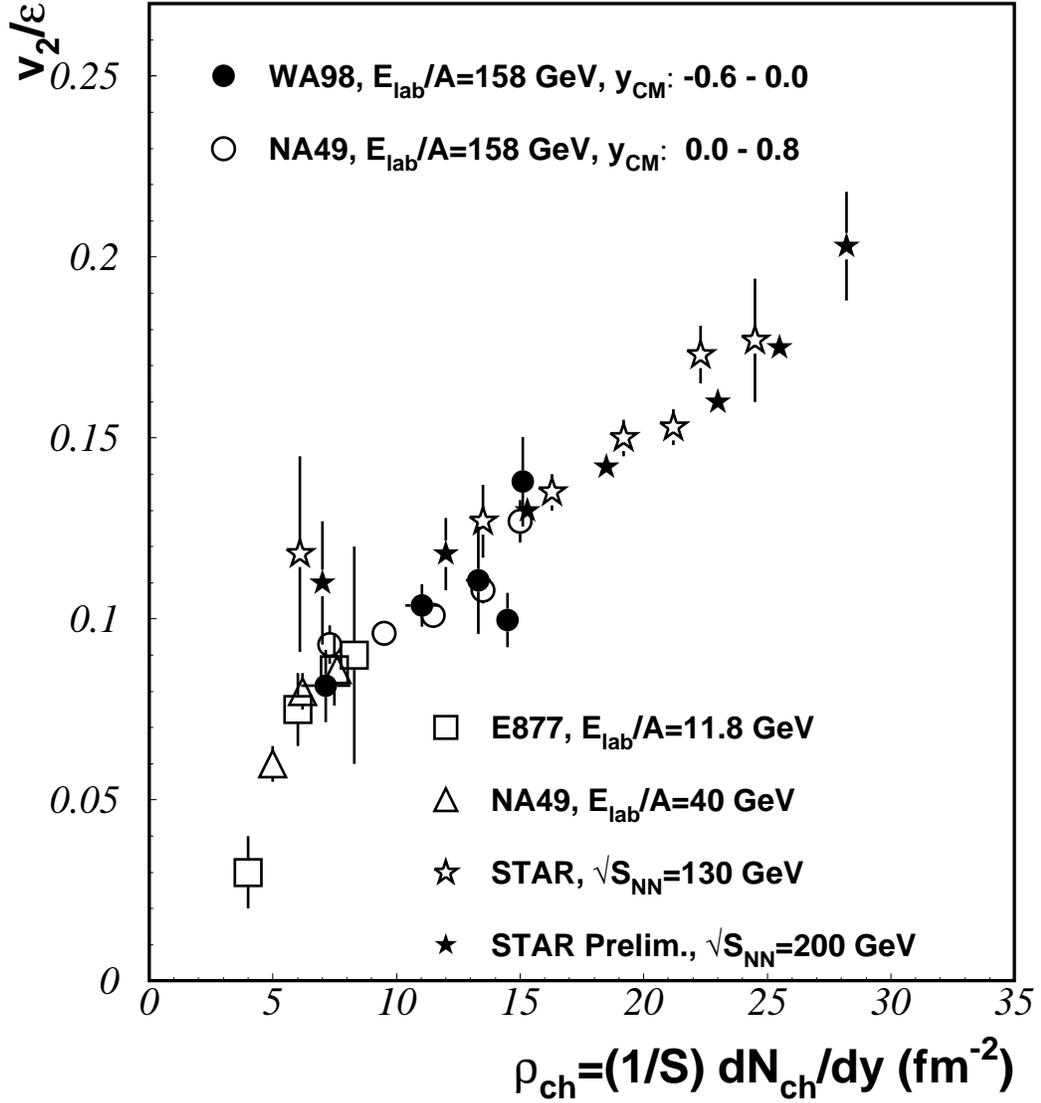}
\caption{\small 
The scaled elliptic flow $v_2/\epsilon$ as a function of
particle density. Results are shown for the WA98 $p_T$-integrated
$v_2$ values at $\sqrt{s_{NN}}=17$ GeV 
( $-0.6 < y_{CM} < 0$).
The WA98 results are compared to results from the E877, NA49, and STAR
experiments.
The WA98 result is the $\pi^0$ elliptic flow extracted from the
measured photon elliptic flow, as described in the text. 
 \protect\label{fig10}}
\end{center}
\end{figure}

\end{document}